\documentclass{styles/svproc}
\usepackage{amsmath,amsfonts,amssymb,bm,mathtools}
\usepackage{array, float, caption}
\pdfoutput=1
\usepackage{graphicx}
\usepackage{color}
\usepackage{url}
\usepackage{ulem}

\newcommand{\pa}{\partial}

\newcommand{\mean}[1]{\langle{#1}\rangle}

\newcommand{\abs}[1]{{|#1|}}

\newcommand{\argmin}{\mathop{\rm arg~min}\limits}

\begin{document}
\mainmatter              
\title{Definition and data-driven reconstruction of asymptotic phase and amplitudes of stochastic oscillators via Koopman operator theory}
\titlerunning{Reconstruction of asymptotic phase and amplitude functions}  
%
\author{Shohei Takata$^{1}$, Yuzuru Kato$^{2}$, and Hiroya Nakao$^{3}$}
\authorrunning{Shohei Takata et al.} 
%
\tocauthor{Shohei Takata, Yuzuru Kato, and Hiroya Nakao}
\institute{
Department of Systems and Control Engineering, Tokyo Institute of Technology, Japan\\
\email{tktsho72@gmail.com}
\and
Department of Complex and Intelligent Systems,
Future University Hakodate, Hokkaido, Japan\\
\email{katoyuzu@fun.ac.jp}
\and
Department of Systems and Control Engineering, Tokyo Institute of Technology, Japan\\
\email{nakao@sc.e.titech.ac.jp}
}
\maketitle              


\begin{abstract}
Asymptotic phase and amplitudes are fundamental concepts in the analysis of limit-cycle oscillators.
In this paper, we briefly review the definition of these quantities,
particularly a generalization to stochastic oscillatory systems from the viewpoint of Koopman operator theory,
and discuss a data-driven approach to estimate the asymptotic phase and amplitude functions
from time-series data of stochastic oscillatory systems.
We demonstrate that the standard Extended dynamic
mode decomposition (EDMD) can successfully reconstruct the phase and amplitude functions of the noisy
FitzHugh-Nagumo neuron model only from the time-series data.
\end{abstract}


\section{Introduction}

Spontaneous rhythmic oscillations and synchronization are widely observed in the real world. Such rhythmic oscillators are typically modeled as nonlinear dynamical systems possessing stable limit-cycle solutions~\cite{Winfree2001geometry,kuramoto1984chemical,pikovsky2001synchronization,ermentrout2010mathematical}. The \textit{asymptotic phase} and  \textit{amplitudes} are fundamental quantities for analyzing and controlling the dynamics of limit-cycle oscillators~\cite{kuramoto1984chemical,shirasaka2017phase,monga2019phase,kuramoto2019concept,shirasaka2020phase,nakao2021phase}.
Recently, it has been shown that the asymptotic phase and amplitudes can be systematically defined based on Koopman operator theory~\cite{mezic2013analysis,mauroy2020koopman} by using the Koopman eigenfunctions of the limit-cycle oscillator~\cite{mauroy2013isostables}, and this definition has further been extended to stochastic oscillators~\cite{thomas2014asymptotic,kato2021asymptotic}. 
In this paper, we briefly review the definition of the asymptotic phase and amplitudes and discuss data-driven estimation of these quantities from time-series data. We apply the standard Extended Dynamic Mode Decomposition (EDMD)~\cite{williams2015data,kutz2016dynamic,klus2020data} to stochastic oscillators, which reconstructs the Koopman eigenvalues and eigenfunctions from time-series data, and show that the asymptotic phase and amplitudes of the noisy 
FitzHugh-Nagumo neuron model can be successfully reproduced.

\section{Asymptotic phase and amplitudes for deterministic limit-cycle oscillators}

\subsection{Asymptotic phase and amplitudes}

First, we provide a brief review of the concepts of asymptotic phase and amplitudes, along with their definitions from the Koopman operator viewpoint.

The asymptotic phase has been a fundamental concept in the analysis of limit-cycle oscillators~\cite{Winfree2001geometry,kuramoto1984chemical,monga2019phase,nakao2016phase}. 
Let us consider a limit-cycle oscillator, i.e., a dynamical system possessing a stable limit-cycle solution described by
\begin{align}
\frac{d{\bm x}(t)}{dt} = {\bm f}({\bm x}(t)),
\label{eq1}
\end{align}
where ${\bm x}(t) \in {\mathbb R}^N$ is the state of the system at time $t$ and ${\bm f} : {\mathbb R}^N \to {\mathbb R}^N$ is a smooth vector field describing the system dynamics. 
We assume that Eq.~\eqref{eq1} has an exponentially stable limit-cycle solution ${\bm x}_0(t)$, satisfying ${\bm x}_0(t+T) = {\bm x}_0(t)$, where $T$ is the period of the limit cycle. We define the frequency as $\omega = 2\pi / T$.
We denote the limit cycle by $\chi = \{ {\bm x}_0(t)\ |\ 0 \leq t < T \}$ and assign a phase $\theta$ ($0 \leq \theta < 2\pi$) to each state ${\bm x}_0(t=\theta/\omega)$ on $\chi$, where $0$ and $2\pi$ are identified.
The linear stability of $\chi$ is characterized by the Floquet exponents $\lambda_m \in {\mathbb C}$ ($m=1, ..., N-1$)~\cite{guckenheimer1983nonlinear}, whose real parts are negative as $\chi$ is assumed to be stable.
We sort the exponents $\lambda_m$ in decreasing order of their real parts.

In the basin of attraction $B \subset {\mathbb R}^N$ of the limit cycle $\chi$, the asymptotic phase function $\Theta : B \to [0, 2\pi)$, which maps a system state ${\bm x} \in B$ to a phase value in $[0, 2\pi)$, is defined such that 
\begin{align}
{\bm f}({\bm x}) \cdot \nabla \Theta({\bm x}) = \omega
\label{eq2}
\end{align}
is satisfied for any ${\bm x} \in B$, where $\cdot$ represents the scalar product of two vectors and $\nabla = \pa / \pa {\bm x}$ is the gradient of a scalar function of ${\bm x}$.
It can then easily be shown that, for any state ${\bm x}(t) \in B$, the phase $\theta(t) = \Theta({\bm x}(t))$ of ${\bm x}(t)$ evolves with time as
\begin{align}
\frac{d \theta(t)}{dt} = \left. \frac{\pa \Theta}{\pa {\bm x}}\right|_{{\bm x} = {\bm x}(t)} \cdot \frac{d{\bm x}(t)}{dt} = {\bm f}({\bm x}(t)) \cdot \nabla \Theta({\bm x}(t)) = \omega
\label{eq3}
\end{align}
by the chain rule of differentiation. Thus, the asymptotic phase $\theta(t) = \Theta({\bm x}(t))$ of the system state ${\bm x}(t)$ always increases with a constant frequency $\omega$ for any trajectory in $B$.
Since all states in $B$ are eventually attracted to the limit cycle $\chi$, all states on the same level set of the asymptotic phase $\Theta$ converge to the same state on $\chi$, because their phase values are always kept equal by Eq.~\eqref{eq3}.
From a geometrical viewpoint, the above definition of the asymptotic phase is equivalent to assigning the same phase value to the set of system states that eventually converge to the same phase on $\chi$.
Such level sets are called {\it isochrons}~\cite{Winfree2001geometry,kuramoto1984chemical,winfree1967biological,guckenheimer1975isochrons}.
The asymptotic phase has played a fundamental role in analyzing the dynamics of limit-cycle oscillators~\cite{Winfree2001geometry,kuramoto1984chemical,ermentrout2010mathematical,monga2019phase,nakao2016phase}.

In a similar manner to the asymptotic phase function, we can also introduce the amplitude functions $R_m : B \to {\mathbb C}$ ($m=1, ..., N-1$), which map a system state ${\bm x} \in B$ to (generally complex) amplitudes such that
\begin{align}
{\bm f}({\bm x}) \cdot \nabla R_m({\bm x}) = \lambda_m R_m({\bm x})
\label{eq4}
\end{align}
holds for any ${\bm x} \in B$, where $\lambda_m \in {\mathbb C}$ are the Floquet exponents of $\chi$.
As we explain in the next subsection, these $R_m({\bm x})$ are the Koopman eigenfunctions of Eq.~\eqref{eq1} associated with the Koopman eigenvalues $\lambda_m$. 
It can then easily be shown that, for an arbitrary state ${\bm x}(t) \in B$, each amplitude $r_m(t) = R_m({\bm x}(t))$ of the state ${\bm x}(t)$ evolves with time as
\begin{align}
\frac{d r_m(t)}{dt} = \left. \frac{\pa R_m({\bm x})}{\pa {\bm x}}\right|_{{\bm x} = {\bm x}(t)} \cdot \frac{d{\bm x}(t)}{dt} = {\bm f}({\bm x}(t)) \cdot \nabla R_m({\bm x}(t)) = \lambda_m r_m(t).
\end{align}
Since any ${\bm x}(t) \in B$ converges to $\chi$, $R_m({\bm x}_0(t)) = 0$ should hold for any ${\bm x}_0(t) \in \chi$.
Thus, the amplitude $r_m(t) = R_m({\bm x}(t))$ characterizes the deviation of the state ${\bm x}(t)$ from $\chi$, which always decays linearly with a constant rate $\lambda_m$ determined by the Floquet exponent for any trajectory.
The level sets of $R_m$ are called {\it isostables}~\cite{mauroy2013isostables}.

The asymptotic phase function $\Theta$ and the amplitude functions $R_m$ define a nonlinear coordinate transformation from $B$ to $[0,2\pi) \times {\mathbb C}^{N-1}$, which gives a globally linearized representation, $\dot{\Theta} = \omega$ and $\dot{R}_m = \lambda R_m$, of the original nonlinear dynamics $\dot{\bm x}={\bm f}({\bm x})$ in $B$. Moreover, by eliminating some of the fast-decaying amplitudes, i.e., assuming $R_m = 0$ for some ranges of $m$ corresponding to sufficiently small $\mbox{Re}\ \lambda_m$, we can choose only the slowly decaying amplitudes and approximately reduce the dimensionality of the system. Applications of the reduced phase-amplitude equations for analyzing and controlling of limit cycles have attracted attention recently~\cite{shirasaka2017phase,monga2019phase,wilson2016isostable,wilson2018greater,mauroy2018global,shirasaka2020phase,takata2021fast}.

\subsection{Koopman eigenvalues and eigenfunctions}

The concepts of asymptotic phase and amplitudes are closely related to Koopman eigenfunctions. Indeed, the amplitude functions have been explicitly defined as the eigenfunctions of the Koopman operator only recently in~\cite{mauroy2013isostables}, despite similar concepts of amplitudes having often been used in the past~\cite{kuramoto1984chemical,kuramoto2019concept,teramae2009stochastic,goldobin2010dynamics,wedgwood2013phase,kotani2020nonlinear}.

We denote the flow of the limit-cycle oscillator described by Eq.~\eqref{eq1} as $S^\tau$, satisfying ${\bm x}(t+\tau) = S^{\tau} {\bm x}(t)$, where $\tau \geq 0$. The Koopman operator $U^\tau$ of Eq.~\eqref{eq1} is then defined as
\begin{align}
(U^\tau g)({\bm x}) = g(S^{\tau} {\bm x})
\end{align}
for ${\bm x} \in B$, where $g : B \to {\mathbb C}$ is an observable of the system that assigns a (generally complex) value $g({\bm x})$ to the system state ${\bm x}$. 
The infinitesimal generator of $U^\tau$ is defined as
\begin{align}
A g({\bm x}) = \lim_{\tau \to +0} \frac{ U^\tau g({\bm x}) - g({\bm x}) }{\tau} 
= \lim_{\tau \to +0} \frac{ g(S^\tau {\bm x}) - g({\bm x}) }{\tau} = {\bm f}({\bm x}) \cdot \nabla g({\bm x}),
\end{align}
where we assumed that $g({\bm x})$ can be expanded in ${\bm x}$ and used that $S^\tau {\bm x} = {\bm x} + {\bm f}({\bm x}) \tau + O(\tau^2)$ and $g(S^\tau {\bm x}) = g({\bm x}) + {\bm f}({\bm x}) \tau \cdot \nabla g({\bm x}) + O(\tau^2)$ for sufficiently small $\tau \geq 0$.
It can easily be shown that $U^{\tau}$ and $A$ are linear operators even if ${\bm f}({\bm x})$ is a nonlinear function of ${\bm x}$~\cite{mezic2013analysis,mauroy2020koopman}.
The eigenvalue ${\tilde{\lambda}} \in {\mathbb C}$ and associated eigenfunction $\phi_{{\tilde{\lambda}}}({\bm x}) : B \to {\mathbb C}$ of $A$ satisfying
\begin{align}
A \phi_{\tilde{\lambda}}({\bm x}) = {\bm f}({\bm x}) \cdot \nabla \phi_{\tilde{\lambda}}({\bm x}) = {\tilde{\lambda}} \phi_{\tilde{\lambda}}({\bm x})
\end{align}
are called the Koopman eigenvalue and Koopman eigenfunction, respectively. 

For an exponentially stable limit cycle, it is known that the set of Koopman eigenvalues includes $i \omega$ and the Floquet exponents $\lambda_m$ ($m=1, ..., N-1$), which are called the principal Koopman eigenvalues, and analytic observables can be expanded by using the associated principal Koopman eigenfunctions~\cite{mauroy2016global}. 
From Eqs.~\eqref{eq2} and~\eqref{eq4}, the complex exponential of the asymptotic phase, $\Phi({\bm x}) = e^{i \Theta({\bm x})}$, and the amplitudes $R_m$ ($m=1, ..., N-1)$ satisfy
\begin{align}
A \Phi({\bm x}) = i \omega \Phi({\bm x}), \quad A R_m({\bm x}) = \lambda_m R_m({\bm x}).
\label{eq9}
\end{align}
Thus, the asymptotic phase $\Theta({\bm x})$ is the argument of the Koopman eigenfunction $\Phi({\bm x})$ with the eigenvalue $i \omega$, i.e., $\Theta({\bm x}) = \mbox{Arg}\ \Phi({\bm x})$, and the amplitudes $R_m({\bm x})$ are nothing but the Koopman eigenfunctions with the principal Koopman eigenvalues $\lambda_m$ given by the Floquet exponents.

Thus, recent developments in Koopman operator theory for dissipative dynamical systems have clarified that the asymptotic phase of the limit cycle, originally introduced from a geometrical viewpoint~\cite{winfree1967biological}, can be defined using the Koopman eigenfunction, and also enabled to define the amplitudes of limit cycle explicitly as eigenfunctions of the Koopman operator in a natural and unified manner. 
This motivates us to define the asymptotic phase and amplitudes for stochastic oscillatory systems also from the Koopman operator viewpoint.

\section{Asymptotic phase and amplitudes for stochastic oscillatory systems}

\subsection{Stochastic oscillators}

In the real world, there exist many examples of stochastic oscillatory systems in which noise plays essential roles in their dynamics.
In this section, we review the definitions of asymptotic phase and amplitudes for stochastic oscillators based on Koopman operator theory that we recently proposed~\cite{kato2021asymptotic}.

We consider a stochastic oscillator described by the following It\^o stochastic differential equation (SDE):
\begin{align}
	d{\bm x}(t) = {\bm A}({\bm x}(t)) dt + {\bm B}({\bm x}(t)) d{\bm W}(t),
	\label{sde}
\end{align}
where ${\bm x}(t) \in {\mathbb R}^{N}$ represents the system state at time $t$, ${\bm A}({\bm x}) \in {\mathbb R}^N$ is a vector field representing the deterministic part of the dynamics, ${\bm B}({\bm x}) \in {\mathbb R}^{N \times N}$ is a matrix representing the noise intensity, and ${\bm W}(t) \in {\mathbb R}^N$ is a $N$-dimensional Wiener process~\cite{gardiner2009stochastic,pavliotis2016stochastic}.
We assume that ${\bm A}({\bm x})$ and ${\bm B}({\bm x})$ satisfy appropriate conditions on smoothness and boundedness~\cite{pavliotis2016stochastic}.
Since we consider stochastic oscillators, we assume that the deterministic part ${\bm A}({\bm x})$ possesses an exponentially stable limit-cycle solution as before.
By the effect of noise characterized by ${\bm B}({\bm x})$, the average properties of the stochastic oscillator, e.g., the period, shape of the trajectory, response properties, can differ from the deterministic case.

As an example, we will consider a stochastic FitzHugh-Nagumo model of a spiking neuron~\cite{kato2021asymptotic} described by the following It\^o SDE:
\begin{align}
	\label{eq:fhz}
	dx &= ( x - a_1x^3 - y ) dt + \sqrt{D_x} dW_x,
	\cr
	dy &= \eta_1 (x + b_1) dt + \sqrt{D_y} dW_y,
\end{align}
where $x$ and $y$ represent real variables,
$a_1$, $b_1$, and $\eta_1$ are parameters of the system, 
$W_x$ and $W_y$ represent independent Wiener processes, 
and $D_x$ and $D_y$ are the intensities of the noise, respectively.
The FitzHugh-Nagumo model is a two-dimensional model of spiking neurons simplified from the more realistic Hodgkin-Huxley model with four variables~\cite{ermentrout2010mathematical}.
We set the parameters as $(a_1, b_1,$ $\eta_1, D_x, D_y)$ $= (1/3, 0.5, 0.5, 0.2, 0.2)$~\cite{kato2021asymptotic}, with which the deterministic part of the system possesses a stable limit-cycle solution.

\begin{figure} [!t]
	\begin{center}
		\includegraphics[width=0.35\hsize,clip]{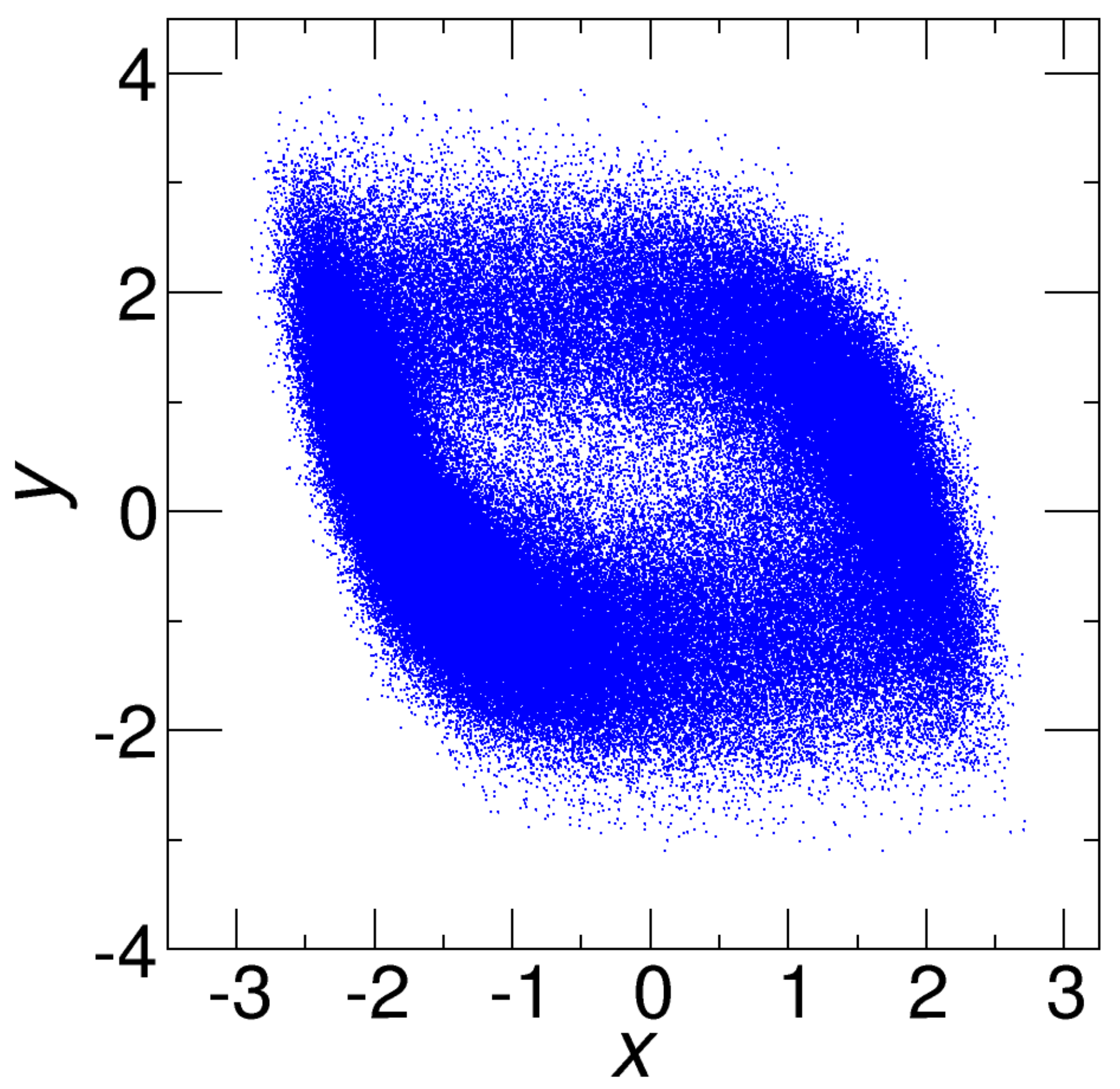}
		\caption{
			Time-series data obtained from the stochastic FitzHugh-Nagumo model.}
		\label{fig1}
	\end{center}
\end{figure}

Figure~\ref{fig1} shows an example of the stochastic trajectory of Eq.~\eqref{eq:fhz}, simulated by the Euler-Maruyama method with a time step $10^{-3}$ and sampled with a time interval $\tau = 0.1$. Due to relatively strong noise, the trajectory of the oscillator shows strong fluctuations.

\subsection{Stochastic Koopman operator}

The Koopman operator for the stochastic oscillator described by Eq.~\eqref{sde} is introduced as follows.
The transition probability density $p({\bm x}, t| {\bm y}, s)$ ($t \geq s$)
of Eq.~(\ref{sde}) obeys the forward and backward Fokker-Planck equations~\cite{gardiner2009stochastic},
\begin{align}
	\frac{\pa}{\pa t} p({\bm x}, t| {\bm y}, s) = {L}_{\bm x}  p({\bm x}, t| {\bm y}, s),
	\quad
	\frac{\pa}{\pa s} p({\bm x}, t| {\bm y}, s) = -{L}^*_{\bm y} p({\bm x}, t| {\bm y}, s),
\end{align}
where the forward  and  backward Fokker-Planck operators are  given by
\begin{align}
	\label{eq:fpe}
	{L}_{\bm x} = - \frac{\pa}{\pa {\bm x}} {\bm A}({\bm x}) + \frac{1}{2} \frac{\pa^2}{\pa {\bm x}^2} {\bm D}({\bm x}),\quad	
	{L}_{\bm x}^* = {\bm A}({\bm x}) \frac{\pa}{\pa {\bm x}} + \frac{1}{2} {\bm D}({\bm x}) \frac{\pa^2}{\pa {\bm x}^2}.
\end{align}
Here, ${\bm D}({\bm x}) = {\bm B}({\bm x}) {\bm B}({\bm x})^{\sf T} \in {\mathbb R}^{N \times N}$ is a matrix of diffusion coefficients with ${\sf T}$ representing the matrix transposition.
The forward and backward operators $L_{\bm x}$ and $L_{\bm x}^*$ are mutually adjoint, i.e., 
$	\mean{ {L}_{{\bm x}} G({\bm x}), H({\bm x})}_{{\bm x}}
=\mean {G({\bm x}), {L}^*_{\bm x} H({\bm x})}_{{\bm x}}$,
where the inner product is defined as
$	\mean{G({\bm x}), H({\bm x})}_{{\bm x}} = \int \overline{ G({\bm x}) } H({\bm x}) d {\bm x}$
for two functions $G({\bm x}), H({\bm x}) : {\mathbb R}^N \to {\mathbb C}$ with the overline indicating complex conjugate, and the integration is taken over the whole range of ${\bm x}$.

The linear differential operators ${L}_{\bm x}$ and ${L}_{\bm x}^*$ have 
a biorthogonal eigensystem $\{\mu_{k}, P_{k}, Q_{k}\}_{k=0, 1, 2, ...}$ of
eigenvalues $\mu_{k}$ and eigenfunctions $P_k({\bm x})$ and $Q_k({\bm x})~(k=0, 1, 2, ...)$,
satisfying
\begin{align}
	\label{eq:eig_fun}
	{L}_{\bm x} P_{k}({\bm x}) = \mu_k P_{k}({\bm x}),
	\quad
	{L}_{\bm x}^* Q^{}_{k}({\bm x}) = \overline{ \mu_k } Q_{k}({\bm x}),
	\quad
	\mean{ P_{k}({\bm x}), Q_{l}({\bm x})}_{{\bm x}} 
	= \delta_{kl},
\end{align}
where $k, l = 0, 1, 2, \ldots$ and $\delta_{kl}$ is the Kronecker delta~\cite{gardiner2009stochastic}.
We sort the eigenvalues $\mu_k$ in decreasing order of their real parts.
Among the eigenvalues, one eigenvalue is zero, $\mu_0=0$, which is associated with the stationary probability density function $P_0({\bm x})$ of the system satisfying $L_{\bm x} P_0({\bm x}) = 0$, and all other eigenvalues have negative real parts.

For Eq.~\eqref{sde}, the stochastic Koopman operator $U_{st}$ for an observable $g:{\mathbb R}^N \to {\mathbb C}$ is defined as~\cite{mezic2005spectral}
\begin{align}
U_{st}^{\tau} g({\bm x}) = {\mathbb E}[ g(S_{st}^{\tau} {\bm x}) ] = \int p({\bm y}, \tau | {\bm x}, 0) g({\bm y}) d{\bm y},
\end{align}
where $S_{st}^\tau$ is a stochastic flow of Eq.~\eqref{sde} and $\mathbb E$ represents the expectation over realizations of $S_{st}^\tau$,
and the infinitesimal generator of $U_{st}^\tau$ is defined as
\begin{align}
A_{st} g({\bm x}) 
= \lim_{\tau \to +0} \frac{U_{st}^{\tau} g({\bm x}) - g({\bm x})}{\tau}
= \lim_{\tau \to +0} \frac{{\mathbb E}[ g(S_{st}^{\tau} {\bm x}) ] - g({\bm x})}{\tau}.
\end{align}

It can be shown that the backward Fokker-Planck operator $L_{\bm x}^*$ in Eq.~(\ref{eq:fpe}) is the infinitesimal generator of the stochastic Koopman operator~\cite{klus2020data,kato2021asymptotic}, i.e.,
\begin{align}
A_{st} = {L}_{\bm x}^* = {\bm A}({\bm x}) \frac{\pa}{\pa {\bm x}} + \frac{1}{2} {\bm D}({\bm x}) \frac{\pa^2}{\pa {\bm x}^2}.
\end{align}
Indeed, from the It\^o formula~\cite{gardiner2009stochastic}, $g({\bm x}(t))$ obeys
\begin{align}
dg({\bm x}(t)) = \left( {\bm A}({\bm x}) \frac{\pa g}{\pa {\bm x}} + \frac{1}{2} {\bm D}({\bm x}) \frac{\pa^2 g}{\pa {\bm x}^2} \right) dt
+ \left( \frac{\pa g}{\pa {\bm x}} \right)^{\top} {\bm B}({\bm x}) d{\bm W}(t),
\end{align}
and, using $U_{st}^{dt} g({\bm x}) = g(S_{st}^{dt} {\bm x}) = g({\bm x}) + dg({\bm x})$, we obtain
\begin{align}
A_{st} g({\bm x}) = \lim_{\tau \to +0} \frac{{\mathbb E}[ g({\bm x}) + dg({\bm x}) ] - g({\bm x})}{\tau}
=
{\bm A}({\bm x}) \frac{\pa g({\bm x})}{\pa {\bm x}} + \frac{1}{2} {\bm D}({\bm x}) \frac{\pa^2 g({\bm x})}{\pa {\bm x}^2}.
\end{align}
We note that ${\mathbb E}$ operates only on $dg({\bm x})$.
If the noise is absent, ${\bm D}({\bm x}) = 0$, the generator $A_{st}$ will take the form ${\bm A}({\bm x}) \frac{\pa}{\pa {\bm x}} = {\bm A}({\bm x}) \cdot \nabla$, which is the Koopman operator $A$ for the deterministic case.

\subsection{Asymptotic phase and amplitudes for stochastic oscillators}

How to define the asymptotic phase for stochastic oscillators has been discussed in several studies~\cite{thomas2014asymptotic,schwabedal2013phase}.
Recently, we proposed general definitions of the asymptotic phase and amplitudes for stochastic oscillators from the Koopman operator viewpoint~\cite{kato2021asymptotic}, where the deterministic definitions in Eq.~\eqref{eq9} using the Koopman eigenfunctions were naturally generalized to stochastic oscillators. 
In~\cite{kato2022definition,kato2023quantum}, we further developed similar ideas for quantum nonlinear oscillators.

As we are considering stochastic oscillators, we assume that the eigenvalues of $L_{\bm x}$ and $L_{\bm x}^*$ with the largest non-zero real part appear as a complex conjugate pair, denoted by $\mu_1$ and $\mu_2 = \overline{\mu_1}$, where the real part $\mbox{Re}\ \mu_1$ characterizes the decay rate and the imaginary part $\Omega = \mbox{Im}\ \overline{\mu_1}$ characterizes the oscillation frequency (note that $\mu_0 = 0$). We may also take $\mbox{Im}\ {\mu_1}$ as $\Omega$, which reverses the direction of the phase.
We focus on this slowest-decaying oscillatory mode of the transition probability density 
and define the system's phase with respect to this mode.
In general, we obtain a branch of complex-conjugate eigenvalues of $L_{\bm x}^*$ in addition to the above slowest-decaying fundamental mode, but we do not consider those modes because their imaginary parts are essentially integer multiples of $\Omega$ and are not independent from the fundamental mode.
For the amplitude, we choose a non-zero eigenvalue $\mu_r$ with the largest real part that is not included in the above branch as the independent, second-slowest decaying mode, which corresponds to the slowest amplitude in the noiseless deterministic limit~\cite{takata2021fast}. We note that $\mu_r$ may be a complex value when $N \geq 3$.

In~\cite{kato2021asymptotic}, focusing on these Koopman eigenvalues $\mu_1$ and $\mu_r$ of $A_{st} = L_{\bm x}^*$, we proposed to use the associated eigenfunction $Q_1({\bm x})$ and $Q_r({\bm x})$ to define the asymptotic phase and the slowest-decaying amplitude of a stochastic oscillator, respectively.
That is, we consider $\mbox{Arg}\ Q_1({\bm x})$ as the asymptotic phase, i.e., $Q_1({\bm x})$ as the complex exponential of the asymptotic phase, and $Q_r({\bm x})$ as the amplitude, in a similar manner to the deterministic case.
We note that we consider only the asymptotic phase and the slowest-decaying amplitude here.

We can then show that the averaged asymptotic phase and the slowest-decaying amplitude of the system at time $t$, given by
\begin{align}
\theta(t) = \mbox{Arg}\ \mathbb{E} [ Q_1({\bm x}) ],
\quad
r(t) = {\mathbb E} [ {Q_r({\bm x})} ],
\end{align}
satisfy
\begin{align}
	\frac{d \theta(t)}{dt} = \Omega,
	\quad
	\frac{d r(t)}{dt} = \mu_r r(t),
	\label{phaseamplitude}
\end{align}
where $\Omega = \mbox{Im}\ \overline{\mu_1}$ and $\mu_r$ characterize the average oscillation frequency and decay rate of the phase and amplitude, respectively.
In deriving the above equations, we used the fact that the eigenfunctions $Q_k({\bm x})$ of $A_{st} = L_{\bm x}^*$ generally satisfy
\begin{align}
\frac{d}{dt} {\mathbb E}[ Q_k(S_{st}^t {\bm x}) ] = \mu_k {\mathbb E}[ Q_k(S_{st}^t {\bm x}) ]
\label{eq18}
\end{align}
for $k=1, 2, 3, ...$, which can easily be shown by taking the time derivative of the expectation
${\mathbb E}[ Q_k(S_{st}^t {\bm x}) ] = \int p({\bm y}, \tau | {\bm x}, 0 ) Q_k({\bm y}) d{\bm y}$~\cite{kato2021asymptotic}.

Thus, we proposed definitions of the asymptotic phase that, on average, increases with a constant frequency $\Omega$ and 
of the amplitude that, on average, decay linearly with a constant decay rates $\mu_r$ with the stochastic evolution of ${\bm x}$, 
for stochastic oscillatory systems. They can be interpreted as natural generalizations of the corresponding definitions for deterministic oscillators.
The Koopman eigenfunctions $Q_1({\bm x})$ and $Q_r({\bm x})$ for the stochastic oscillator play essentially similar roles to the Koopman eigenfunctions $\phi_0({\bm x})$ and $\phi_1({\bm x})$ for the deterministic oscillator.
Our definition of the asymptotic phase and amplitudes based on the Koopman eigenfunctions coincide with the definition based on the first-passage time problem
by Thomas and Lindner~\cite{thomas2014asymptotic} and  P{\'e}rez-Cervera {\it et al.}~\cite{perez2021isostables} , respectively. Recently, P{\'e}rez-Cervera {\it et al.}~\cite{perez2023universal} also introduced a description of stochastic oscillators using a complex Koopman eigenfunctions $Q_1({\bm x})$. 

\section{Data-driven estimation of the asymptotic phase and amplitude functions}

\subsection{Extended dynamic mode decomposition}

In this section, we aim to reconstruct the asymptotic phase and amplitude functions in a data-driven manner using the Extended Dynamic Mode Decomposition (EDMD), which is a standard method for obtaining the leading Koopman eigenvalues and eigenfunctions from time-series data~\cite{williams2015data,kutz2016dynamic,klus2020data,korda2018convergence}.

First, we briefly review the algorithm of EDMD.
We consider a dictionary consisting of $N_k$ basis functions (observables) $\psi_{1},\ ...,\ \psi_{N_k} : \mathbb{R}^{N} \to {\mathbb R}$ and arrange them in the vector form as
\begin{align}
	{\bm \psi}({\bm x}) = [\psi_1({\bm x}), \psi_2({\bm x}), \cdots, \psi_{N_k}({\bm x})]^{\sf T} \in {\mathbb R}^{N_k \times 1}.
\end{align}
The action of the stochastic Koopman operator $U_{st}^\tau$ on ${\bm \psi}({\bm x})$ is
\begin{align}
	U_{st}^{\tau} {\bm \psi}({\bm x}) 
	= {\mathbb E}[ {\bm \psi}(S_{st}^{\tau} {\bm x}) ].
\end{align}
We seek the optimal matrix $K_{st} \in {\mathbb R}^{N_k \times N_k}$ that gives the best finite-dimensional approximation to $U_{st}^{\tau} {\bm \psi}({\bm x})$, i.e.,
\begin{align}
	U_{st}^{\tau} {\bm \psi}({\bm x}) 
	\approx K_{st} {\bm \psi}({\bm x}).
	\label{finiteK}
\end{align}
The approximation error of the above equation is
${\mathbb E}[ {\bm \psi}(S_{st}^{\tau} {\bm x}) ] - K_{st} {\bm \psi}({\bm x}) =$
${\mathbb E}[ {\bm \psi}(S_{st}^{\tau} {\bm x}) - K_{st} {\bm \psi}({\bm x}) ]$, where ${\mathbb E}$ operates only on $S_{st}^\tau$. 

To find the matrix $K_{st}$ from observed time-series data, we introduce two matrices $Y, Y'\in {\mathbb R}^{N_k \times {M-1}}$ consisting of $M-1$ data points sampled with an interval of $\tau > 0$,
\begin{align}
	\label{eq:daty}
	Y = [{\bm \psi}({\bm x}_1), {\bm \psi}({\bm x}_2),\cdots, {\bm \psi}({\bm x}_{M-1})],
	\quad
	Y' = [{\bm \psi}({\bm x}_2), {\bm \psi}({\bm x}_3),\cdots, {\bm \psi}({\bm x}_{M})],
\end{align}
where ${\bm x}_k = {\bm x}(k \tau)$.
Since ${\bm x}_{k+1} = S_{st}^{\tau} {\bm x}_k$, the mean-squared approximation error in Eq.~(\ref{finiteK}) can be estimated from the time-series data as
\begin{align}
	{\mathbb E} \left[ \|  {\bm \psi}(S_{st}^{\tau}  {\bm x}) - K_{st} {\bm \psi}({\bm x}) \|^2 \right] 
	&
\approx
	\frac{1}{M-1} \sum_{k=1}^{M-1} \|  {\bm \psi}({\bm x}_{k+1}) - K_{st} {\bm \psi}({\bm x}_k) \|^2
	\cr
	&
=
	\frac{1}{M-1} \| Y' - K_{st} Y \|_F^2,
\end{align}
where the Frobenius norm of a matrix $C \in {\mathbb R}^{N_k \times (M-1)}$ is defined as $\| C \|_{F}=\sqrt{\sum_{j=1}^{N_k} \sum_{k=1}^{M-1} C_{j k}^{2}}$.
Thus, the optimal matrix $K_{st}$ is obtained as
\begin{align}
	K_{st} = \argmin_{\tilde{K}}  \left\| {Y}^{\prime} - \tilde{K} {Y} \right\|_{F}^2,
	\label{argminK}
\end{align}
which is explicitly given by
\begin{align}
	\label{eq:B}
	K_{st} 
	= {Y}' {Y}^+ .
\end{align}
Here, ${Y}^+ = Y^{\sf T} ( YY^{\sf T} )^{-1}$ represents the Moore-Penrose pseudoinverse of $Y$.
In practice, we also add a small regularization term proportional to $\| K \|_F^2$ in Eq.~(\ref{argminK}) to avoid overfitting  and obtain the optimal matrix $K$ by 
\begin{align}
	K_{st} = \argmin_{\tilde{K}}  \left\| {Y}^{\prime} - \tilde{K} {Y} \right\|_{F}^2 + \alpha  \left\|  \tilde{K}  \right\|_{F}^2,
	\label{argminK2}
\end{align}
where $\alpha$ is the regularization parameter~\cite{li2017extended}. 
We note that ${\bm x}_{k+1} = S_{st}^{\tau} {\bm x}_k$ represents the stochastic evolution of ${\bm x}_k$ determined by the It\^o SDE~\eqref{sde}. Thus, $K_{st}$ estimated by Eq.~(\ref{argminK2}) corresponds to the stochastic Koopman operator $U_{st}^{\tau}$ and the infinitesimal generator $A_{st}$, i.e., the backward Fokker-Planck operator $L_x^*$.

The approximate Koopman eigenfunctions can be obtained as follows. Using the basis functions ${\bm \psi}({\bm x})$, we approximately represent the Koopman eigenfunction $Q_k({\bm x})$ associated with the eigenvalue $\Lambda_k$ as 
\begin{align}
	Q_k({\bm x}) \approx \sum_{j=1}^{N_k} w^{(k)}_j \psi_j({\bm x}) = {\bm w}_k {\bm \psi}({\bm x}),
	\label{Qkexpansion}
\end{align}
where ${\bm w}_k = [ w^{(k)}_1, w^{(k)}_2, \cdots, w^{(k)}_{N_k} ] \in {\mathbb R}^{1 \times N_k}$ represents the expansion coefficients.
Plugging the above expansion into the eigenvalue equation $U_{st}^{\tau} Q_k({\bm x}) = \Lambda_k Q_k({\bm x})$ for the stochastic Koopman operator, we obtain
\begin{align}
U_{st}^{\tau} Q_k({\bm x}) \approx U_{st}^{\tau} \{ {\bm w}_k {\bm \psi}({\bm x}) \} 
\approx {\bm w}_k K_{st} {\bm \psi}({\bm x})
\approx \Lambda_k {\bm w}_k {\bm \psi}({\bm x}).
\end{align}
Thus, we can approximately obtain $\Lambda_k$ and ${\bm w}_k$ from the eigenvalue equation for $K_{st}$, 
\begin{align}
	{\bm w}_k K_{st} 
	= {\Lambda_k} {\bm w}_k.
\end{align}
The eigenvalue ${\Lambda_k}$ approximates the eigenvalue of $U_{st}^{\tau}$, from which we approximately obtain the Koopman eigenvalue ${\mu_k} = (1/\tau) \ln {\Lambda_k}$ of $A_{st} = L_{\bm x}^*$ and the associated Koopman eigenfunction by Eq.~\eqref{Qkexpansion}.

Thus, the EDMD gives a finite-dimensional approximation to the Koopman operator, eigenvalues, and eigenfunctions projected on the basis ${\bm \psi}({\bm x})$. 
The estimated eigenvalue $\overline{\mu_1}$ gives the frequency $\Omega$ of the stochastic oscillation, the eigenvalue $\mu_r$ gives the decay rate of the slowest amplitude, and the associated eigenfunction $Q_1({\bm x})$ and $Q_r({\bm x})$ give the estimates of the asymptotic phase function and amplitude function, respectively.

\begin{figure} [!t]
	\begin{center}
		\includegraphics[width=1\hsize,clip]{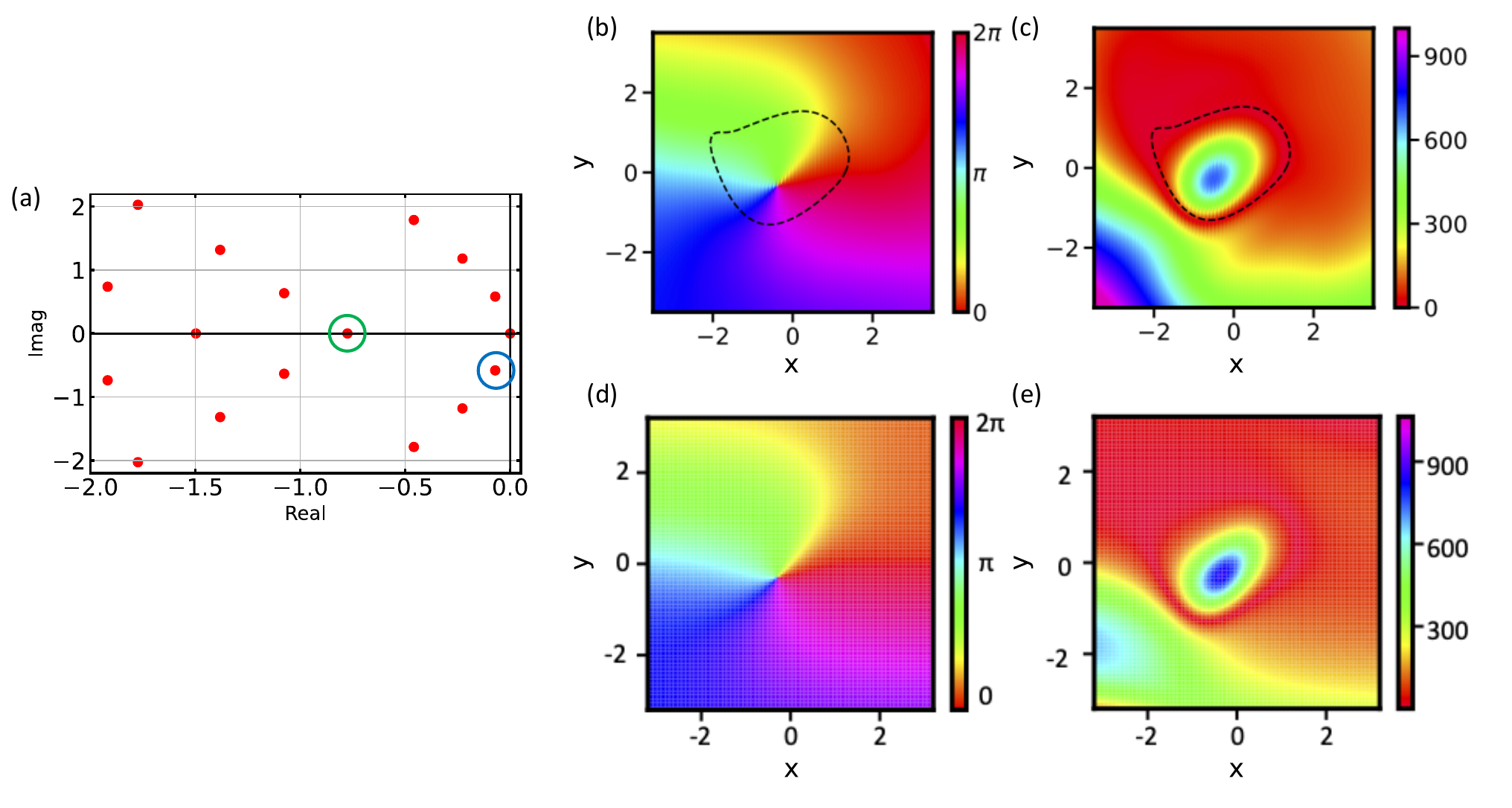}
		\caption{
			Reconstruction of the asymptotic phase and the slowest-decaying amplitude functions of the stochastic FitzHugh-Nagumo model.
			(a) Eigenvalues of $A_{st}$.
			(b) Phase function $\mbox{Arg}\ Q_1({\bm x})$, and (c) modulus of the amplitude function $\abs{Q_r({\bm x})}$ reconstructed by EDMD.
			The zero-level set of the amplitude function $Q_r({\bm x})$ representing the stochastic periodic orbit of the system
			is drawn by the black-dotted line.
			(d) Phase function $\mbox{Arg}\ Q_1({\bm x})$ and (e) modulus of the amplitude function $\abs{Q_r({\bm x})}$ obtained from the Koopman operator~\cite{kato2021asymptotic}.
		}
		\label{fig_1}
	\end{center}
\end{figure}

\subsection{Example: noisy FitzHugh-Nagumo oscillator}

As an example, we consider a stochastic FitzHugh-Nagumo model, Eq.~\eqref{eq:fhz}, and reconstruct the asymptotic phase and amplitude functions, i.e., $\mbox{Arg}\ Q_1({\bm x})$ and $Q_r({\bm x})$, from time-series data by EDMD.
We observed a single trajectory of Eq.~(\ref{eq:fhz}) with a sampling interval of $\tau = 10^{-1}$ (see Fig.~\ref{fig1}).
Following~\cite{williams2015data}, we used $1,000$ radial basis functions (RBFs) $\psi_j({\bm x}) = (r -c_j)^2 \ln (\abs{r - c_j} +\varepsilon)~(j = 1, \ldots, 1000)$ as the observation functions, where the centers $c_j$ are determined by $k$-mean clustering of the sampled data and $\epsilon = 10^{-4}$, to construct the two data matrices in Eq.~(\ref{eq:daty}) (see~\cite{williams2015data} for the details of the RBFs). The regularization parameter in Eq.~\eqref{argminK2} was set as $\alpha = 10^{-3}$.
In the numerical analysis, we confirmed that $10^6$ data points were enough for reconstructing the phase and amplitude functions with appropriate accuracy and that reasonably accurate reconstruction was also possible with only $10^5$ data points. 

Figure~\ref{fig_1}(a) shows the eigenvalues of the infinitesimal generator of the Koopman operator $A_{st} = L^{*}_{\bm x}$ estimated from the time series, where $\mu_1$ and $\mu_r$ are indicated by blue and green circles, respectively. We can observe that the Koopman eigenvalues with the largest non-zero real parts are mutually complex conjugate as assumed, which correspond to $\mu_1$ and $\overline{\mu_1}$, respectively. We choose the real eigenvalue $\mu_r$ belonging to the next branch for the slowest-decaying amplitude.
The average frequency $\Omega$ of the phase is given by the imaginary part of $\overline{\mu_1}$ and estimated as $\Omega \approx 0.581$, 
and the decay rate of the amplitude $\mu_r$ is given by the largest negative real eigenvalue and estimated as $\mu_r \approx -0.775$. 
The reconstructed phase and amplitude functions are shown in Fig.~\ref{fig_1}(b) and (c).
In these figures, the zero-level set of  the amplitude function is shown by a black-dotted curve, which can be regarded as an effective
periodic trajectory of the stochastic oscillator.

For comparison, the frequency and decay rate obtained by direct numerical evaluation of the eigenvalues and eigenfunctions of $L_{\bm x}^*$ are $\Omega \approx 0.582$ and $\mu_r \approx -0.778$~\cite{kato2021asymptotic},
and the associated asymptotic phase and amplitude functions
are  plotted in  Fig.~\ref{fig_1}(d) and (e).
We can confirm that the results obtained from the time-series data by EDMD agree well with the results evaluated directly from $L_{\bm x}^*$.
Thus, EDMD could successfully reconstruct the phase and amplitude functions from time-series data in this example.

\section{Summary}

We have briefly reviewed the concepts of asymptotic phase and amplitudes of deterministic limit-cycle oscillators and demonstrated their natural generalization to stochastic oscillators from the viewpoint of Koopman operator theory. We then discussed the data-driven reconstruction of the asymptotic phase and amplitude functions from time-series data of stochastic oscillators, illustrating that the standard Extended Dynamic Mode Decomposition (EDMD) algorithm for estimating Koopman eigenvalues and eigenfunctions could successfully reconstruct those functions for an example of the stochastic FitzHugh-Nagumo model with reasonable accuracy.
In this article, we focused solely on the definition and estimation of the phase and amplitude functions. In controlling stochastic oscillators based on the estimated functions, such as synchronizing them with periodic input signals, it is also crucial to characterize the effect of external perturbations on the phase and amplitudes of the stochastic oscillators. Further research will be required on this issue.

\textit{---Acknowledgments}
We acknowledge 
JSPS KAKENHI 22K14274, 22K11919, 22H00516, JST CREST JP-MJCR1913 and JST PRESTO JPMJPR24K3 for financial support. 



\end{document}